\documentclass[aps,prb,twocolumn,groupedaddress,showpacs]{revtex4-1}

\usepackage{graphicx}
\usepackage{dcolumn}
\usepackage{bm}
\usepackage{amsmath,amssymb}

\def\rnum#1{\expandafter{%
\romannumeral #1}}
\def\Rnum#1{\uppercase\expandafter{%
\romannumeral #1}}

\newcommand{\bol}[1]{\boldsymbol #1}

\begin{document}


\title{Spin-Chirality Separation and ${\bol S}_3$-Symmetry Breakings 
in the Magnetization Plateau of the Quantum Spin Tube
}
\author{Kouichi Okunishi,$^{1}$ 
Masahiro Sato,$^{2}$ 
T\^oru Sakai,$^{3,4}$
Kiyomi Okamoto,$^{5}$ and 
Chigak Itoi$^{6}$ }
\affiliation{
$^1$ Department of Physics, Niigata University, Niigata 950-2181, Japan\\
$^2$Department of Physics and Mathematics, Aoyama Gakuin University, Sagamihara, Kanagawa 252-5258, Japan\\
$^3$ Japan Atomic Energy Agency, SPring-8, Sayo, Hyogo 679-5148, Japan\\
$^4$ Department of Material Science, University of Hyogo, Kamigori, Hyogo 678-1297, Japan\\
$^5$ Department of Physics, Tokyo Institute of Technology, Meguro-ku, Tokyo 152-8551, Japan\\
$^6$Department of Physics, Nihon University, Kanda-Surugadai, Chiyoda-ku, Tokyo 101-8308, Japan
}

\date{\today}

\begin{abstract}
We study the magnetization plateau state of the three-leg spin-$\frac{1}{2}$ 
tube in the strong rung coupling region, where $S_3$-symmetry breakings and 
the low-energy chirality degree of freedom play crucial roles. 
On the basis of the effective chirality model and density matrix renormalization group, we clarify that, as the leg coupling increases, the chirality liquid with gapless non-magnetic excitations, the spin imbalance phase and the vector-spin-chirality ordered phase emerge without closing the plateau spin gap. 
The relevance of these results to experiments is also discussed.
\end{abstract}

\pacs{75.10.Jm, 75.10.Pq, 75.30.Kz, 75.40.Cx}

\maketitle

\section{Introduction}
Geometrical frustration on magnetism has long been one of the attractive 
subjects in condensed-matter and statistical physics, 
since the frustration provides rich physical phenomena and various 
ordered/disordered states \cite{frustration}. 
It is well-established that the spin chirality often plays a fundamental 
role  as we probe the frustration effects, especially, in the triangular lattice systems~\cite{Miyashita,chirality}.
Recently multiple-spin orders without any magnetic moment, 
including vector spin chiral order, have been actively studied as a new topic 
in frustrated magnetism (e.g., one- and two-dimensional $J_1$-$J_2$ 
spin models~\cite{Okunishi08,Shannon,HKMF,Furukawa}).
The vector spin chirality also attracts extensive attention in the context 
of multiferroics~\cite{Wang}, where the chirality order 
induces electric polarization.
In the most of frustrated systems like the $J_1$-$J_2$ models, however, 
the chirality excitation is usually embedded in conventional 
magnetic excitations, which make direct observation of the chirality difficult.
In order to gain deeper understanding of the frustration physics,  thus, it may be a key issue to extract the chirality excitation energetically separated from the magnetic fluctuations in a realistic situation.

\begin{figure}[t]
\begin{center}
\includegraphics[width=7cm]{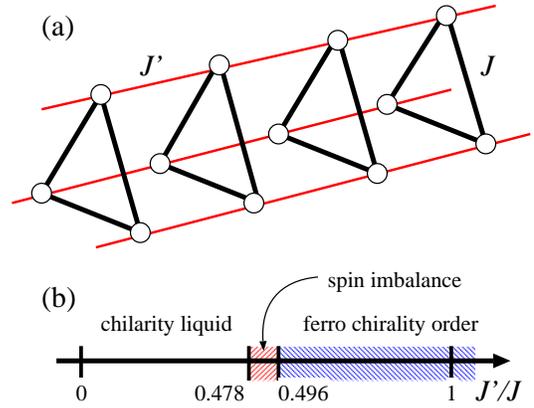}
\end{center}
\caption{(color online) (a) Structure of the three-leg spin tube and (b) 
ground-state phase diagram of the $\frac{1}{3}$ plateau state 
of the spin tube. The plateau is predicted to vanish at a strong 
leg-coupling point $J/J'\sim 0.1$~\cite{Sato07,cabra} (see the text).}
\label{fig:Phases}
\end{figure}

Among a mount of frustrating systems, the three-leg spin tube, 
consisting of coupled three spin-$\frac{1}{2}$ antiferromagnetic chains 
[see Fig.~\ref{fig:Phases} (a)], is one of the models deeply related 
to spin chirality; We can define clockwise/anticlockwise rotation 
along the rung in the spin tube. 
In fact, the topological structure of the spin tube is known to induce 
several interesting 
phenomena~\cite{schulz,kawano,cabra,Sato07,SatoSakai07,mila,fouet,okunishi,sakai,arikawa,review,penc}. 
Recently, spin-tube materials such as 
[(CuCl$_2$tachH)$_3$Cl]Cl$_2$~\cite{Nojiri,Nojiri2} and 
$\rm CsCrF_4$~\cite{Manaka,Manaka2} have been really synthesized and 
characteristic properties to the spin tube have been revealed by several 
experimental approaches. In particular, it is pointed out that 
the broad peak of specific heat is associated with a gapful chirality 
excitation in the {\it twisted} tube 
[(CuCl$_2$tachH)$_3$Cl]Cl$_2$~\cite{Nojiri2}.
However, it should be also noted that the contribution from gapless 
magnetic excitation overlaps this broad peak related to chirality.

In this paper, we demonstrate that the quantum phase transitions 
associated with the chirality actually occur in the magnetization 
plateau of the {\it straight} quantum spin tube, 
where energy scale of the chirality is certainly separated from gapful 
magnetic excitations. The Hamiltonian of the spin tube is given by
\begin{eqnarray}
{\cal H} = \sum_{i=1}^3 \sum_{j=1}^{L}[ J {\bol S}_{i,j}{\bol S}_{i+1,j}  
+J' {\bol S}_{i,j}{\bol S}_{i,j+1} ] 
- H \sum_{i,j} S_{i,j}^z, \nonumber\\
\label{tube}
\end{eqnarray}
where $\bol S_{i,j}$ is the spin-$\frac{1}{2}$ matrix, $J (J')>0$ is the 
intra(inter)-triangle coupling, and $i$ ($j$) represents the label of 
the rung (leg) direction ($i$: mod 3).
This model (\ref{tube}) looks very simple, but the frustration due to 
the tube structure is expected to induce various characteristic properties. 
In fact, it was shown that the model~(\ref{tube}) has a uniform vector spin 
chirality order in the weak rung-coupling region ($J\ll J'$) 
in a magnetic field $H$~\cite{Sato07,SatoSakai07}. 
A rather interesting parameter region is the strong-coupling limit ($J\gg J'$), where the system is basically described by the weakly coupled triangles. 
In the strong rung limit, the composite spin 
\begin{eqnarray}
{\bol T}_j &=& {\bol S}_{1,j}+{\bol S}_{2,j}+{\bol S}_{3,j}
\end{eqnarray} 
on each unit triangle is classified into 
$T=\frac{3}{2} \oplus \frac{1}{2} \oplus \frac{1}{2}$ sectors and then the 
$T^z=\frac{1}{2}$ states of $T=\frac{1}{2}$ sectors lead to a robust 
magnetization plateau at $\frac{1}{3}$ of the full moment~\cite{cabra}. 
A key point is that the two-fold degeneracy of $T=\frac{1}{2}$ sectors in this 
plateau state brings an active low-energy variable, which is just 
the chirality degree of freedom. 
Utilizing the low-energy effective model and density matrix renormalization 
group (DMRG), we will show that the energetic separation of the spin and 
chirality excitations leads to nontrivial quantum phase transitions 
without destroying the magnetization plateau. 
The main results are summarized in Fig.~\ref{fig:Phases} (b); we find 
chirality liquid, spin imbalance, and the ferro-chirality ordered phases. 
We also explain that these orders are accompanied by the 
$S_3$-symmetry breaking in the unit triangle.

The remaining part of this paper is organized as follows. 
In Sec.~\ref{Sec2}, we study the $\frac{1}{3}$ plateau state 
based on the effective spin chirality model. 
We also discuss the role of the $S_3$-symmetry in the quantum spin tube. 
Section~\ref{Sec3} is devoted to the numerical results derived from 
DMRG method. 
Combining the DMRG results with the analytical predictions in Sec.~\ref{Sec2}, 
we reveal three new phases in the plateau region; chirality 
liquid, spin-imbalance, and the ferro-chirality ordered phases. 
Finally we summarize our result and the relation between it and 
previous studies in Sec.~\ref{Sec4}.
Furthermore, we discuss the relevance of our result to experiments.

\section{effective chirality model and $S_3$ symmetry}
\label{Sec2}
Let us start with the low-energy effective theory for the plateau state 
in the strong rung-coupling region. 
We can represent the two-fold degenerating bases for 
the $T^z=1/2$ states of $T=\frac{1}{2}$ on each triangle as 
\begin{subequations}
\begin{eqnarray}
|L\rangle = (|\downarrow \uparrow \uparrow\rangle 
+\omega^{} |\uparrow \downarrow \uparrow\rangle 
+\omega^{-1}|\uparrow \uparrow \downarrow\rangle )/\sqrt{3},\\
|R\rangle =  (|\downarrow \uparrow \uparrow\rangle 
+\omega^{-1} |\uparrow \downarrow \uparrow\rangle 
+\omega^{}|\uparrow \uparrow \downarrow\rangle )/\sqrt{3},
\label{basis}
\end{eqnarray}
\end{subequations}
where $\omega=e^{2\pi i/ 3}$ and $L$ $(R)$ denotes the 
left- (right-) handed mode in the rung direction \cite{schulz}.
These two states indeed stand for the chirality degree of freedom.
By projecting out the high energy states with $T^z=-\frac{1}{2}$ and $T=\frac{3}{2}$  in every unit triangle, the effective Hamiltonian of 
the plateau state is obtained as 
\begin{eqnarray}
{\cal H}_{\rm eff} &=& \sum_j \Big[
\frac{ K_{xy} }{2} (\tau_j^+\tau_{j+1}^- + \tau_j^- \tau_{j+1}^+) 
+   K_z \tau_j^z \tau_{j+1}^z  \nonumber \\  
&+& \frac{K'_{xy}}{2} ( \tau_{j-1}^+\tau_{j+1}^- + \tau_{j-1}^- \tau_{j+1}^+)  
\nonumber \\
&+& \frac{K_3}{4}( \tau_{j-1}^+\tau_j^+\tau_{j+1}^+ 
+ \tau_{j-1}^-\tau_j^-\tau_{j+1}^-) \Big] ,
\label{effective}
\end{eqnarray}
where ${\bol \tau}_j$ is the pseudo-spin-$\frac{1}{2}$ matrix defined by 
$\tau_j^z=(|L\rangle_j{}_j\langle L|-|R\rangle_j{}_j\langle R|)/2$.
The coupling constants are evaluated as $K_{xy}=2J'/3-5J'^2/(27J)$, $K_z=-J'^2/J$, $K'_{xy}=8J'^2/(27J)$ and $K_3= -16J'^2/(27J)$ within the second-order 
perturbation in $J'$.
Here, it is worthy to note that the relation  between ${\bol \tau}_j$ and ${\bol S}_{i,j}$ is given by $\tau^z_j = \sqrt{3} \hat{P}_j\chi_j\hat{P}_j$ and  $\tau^x_j=-\hat{P}_j\mu_j\hat{P}_j$, where
\begin{subequations}
\begin{eqnarray}
 \chi_j&=&\sum_{i=1}^3({\bol S}_{i,j}\times {\bol S}_{i+1,j})^z/3,\\
\mu_j&=& S^z_{1,j}-(S^z_{2,j}+S^z_{3,j})/2,
\end{eqnarray}
\end{subequations}
are respectively the $z$ component of the vector spin chirality and 
an imbalanced magnetization on each triangle, and 
$\hat{P}_j=|L\rangle_j{}_j\langle L|+|R\rangle_j{}_j\langle R|$ 
is the projection operator to the $T^z_j=\frac{1}{2}$ states of $T=\frac{1}{2}$. 

In order to resolve possible quantum phase transitions, 
it is very instructive to discuss the discrete symmetry of the spin tube.
The spin tube has $S_3$-group ($\cong C_{3v}$ point group) symmetry in the rung direction in addition to the translational symmetry along the leg direction.  
The operations in the $S_3$ group are composed of the cyclic permutation 
${\bol S}_{i,j} \to {\bol S}_{i+1,j}$ with mod 3 and the reflection 
${\bol S}_{i,j} \leftrightarrow {\bol S}_{i',j}$ 
at a bond in every unit triangle ($i\neq i'$).
Possible $S_3$-symmetry breakings are classified by its subgroups:
(a) the bond-parity breaking with conserving the cyclic symmetry, 
(b) the cyclic $Z_3$ symmetry breaking with conserving a part of bond-party symmetry, or 
(c) the full breaking of the $S_3$ symmetry. 
The vector spin chirality $\chi_j$ is a typical order parameter in the case (a), which changes its sign by the reflection, but is invariant under the cyclic permutation. 
This cyclic symmetry is related to the spin current circulating in the rung direction.
On the other hand, $\mu_j$ can be an order parameter of the case (b), 
since its form changes via the cyclic permutation, but is invariant under 
the reflection ${\bol S}_{2,j} \leftrightarrow {\bol S}_{3,j}$. 
If $\mu_j$ becomes finite, it suggests that the isosceles-triangle-type 
imbalance occurs for $\langle S^z_{i,j}\rangle $ in the plateau state.

We discuss the relation between the $S_3$ symmetry and the effective model 
(\ref{effective}). 
Write the cyclic permutation operation of the $S_3$ symmetry group as 
${\cal T}_c$, and  the bond reflection as ${\cal T}_r(={\cal T}_r^{-1})$. 
In the level of the effective chirality $\bol \tau$, the $S_3$ symmetric 
operation is given by
\begin{eqnarray}
&&{\cal T}_c \tau^z_j  {\cal T}_c^{-1} =\tau^z_j, 
\quad {\cal T}_c \tau_j^+{\cal T}_c^{-1}= \omega \tau_j^+, 
\quad  {\cal T}_c \tau_j^+{\cal T}_c^{-1}= \omega^2 \tau_j^+,\nonumber \\
&&{\cal T}_r \tau_j^z  {\cal T}_r =-\tau_j^z, 
\quad {\cal T}_r \tau_j^+{\cal T}_r= \tau_j^- , 
\quad {\cal T}_r \tau_j^-{\cal T}_r= \tau_j^+ 
\end{eqnarray}
for any $j$.
Under these operations of the $S_3$ symmetry, the effective Hamiltonian (\ref{effective}) is confirmed to be invariant.
Here we should remark that in the model~(\ref{effective}), the second-roder perturbation process generates the U(1)-symmetry breaking $K_3$ term, although the U(1)-symmetric XY model, which is obtained within the first-order perturbation, has been often used for the spin tubes~\cite{kawano,fouet,mila}. 
This is consistent with the fact that $\tau^z_j\sim\chi_j$ is not exactly conserved in the original spin tube. 
Thus we need a careful consideration about the role of symmetry and 
interactions in the effective model~(\ref{effective}).

According to the bosonization approach~\cite{Giamarchi}, 
the low-energy physics of the model (\ref{effective}) is described 
by a massless free boson theory with several interactions. 
The effective Hamiltonian for the free boson, 
i.e., the Tomonaga-Luttinger (TL) liquid is represented as 
\begin{eqnarray}
{\cal H}_{\rm TL}&=&\int dx \frac{v}{2}\left[\tilde{K}(\partial_x\theta)^2+
\tilde{K}^{-1}(\partial_x\phi)^2\right],
\label{TLL}
\end{eqnarray}
where $(\phi,\theta)$ is the canonical pair of scalar fields 
($x=ja$ and $a$ is lattice spacing), 
$\tilde{K}$ is the TL-liquid parameter, and $v$ is the low-energy excitation 
velocity of the model~(\ref{effective}). 
The effective spin ${\bol \tau}_j$ and the bosonic fields $(\phi,\theta)$ 
is related as 
\begin{eqnarray}
\tau^z_j \simeq  \frac{a}{\sqrt{\pi} }\frac{ \partial \phi(x)}{\partial x} 
+  (-)^j a_1  \cos \sqrt{4\pi} \phi(x) +\cdots, \nonumber \\
\tau^+_j \simeq  e^{i \sqrt{\pi} \theta(x) }
[(-)^j b_0 +  b_1 \cos \sqrt{4\pi }\phi(x) +\cdots],
\end{eqnarray}
with non-universal constants $a_1$, $b_0$ and $b_1$.
The $S_3$-symmetry operations on the effective fields are summarized as 
\begin{eqnarray}
&{\cal T}_r \theta (x) {\cal T}_r= -\theta(x), \hspace{0.5cm} 
{\cal T}_r \phi (x) {\cal T}_r= -\phi(x) +\sqrt{\pi}/2,\nonumber\\
&{\cal T}_c \theta (x) {\cal T}_c^{-1}= \theta(x) + 2 \sqrt{\pi}/3.
\end{eqnarray}
In addition, the operation of one-site translation along the leg 
${\cal T}_l$ transforms the boson fields as 
\begin{eqnarray}
{\cal T}_l \theta (x) {\cal T}_l^{-1}= \theta(x+a)+\sqrt{\pi}, \nonumber\\
{\cal T}_l \phi (x) {\cal T}_l^{-1}= \phi(x+a) +\sqrt{\pi}/2.
\end{eqnarray}
These symmetries impose significant restriction to the possible 
interaction terms in the effective field theory. Among various vertex 
operators permitted by the $S_3$ and translational symmetries, 
the most relevant terms are given by for $\cos (2\sqrt{2\pi} \phi )$ and 
$\cos(6\sqrt{\pi} \theta) $, for which the scaling dimensions are 
respectively  $4\tilde{K}$ and $9/\tilde{K}$. Since the value of $\tilde{K}$ approaches 
unity in the $J'/J \to 0$ limit (the XY model), we can see that 
the interaction terms in Eq. (\ref{effective}) are all irrelevant 
for sufficiently small $J'$, suggesting that the critical chirality liquid 
is realized in a certain region of small $J'$. 
On the other hand, the system may have two kind of instabilities 
as $J'$ increases. 
The first case is the ferro-chirality order of $\tau^z_j\sim\chi_j$. 
Since the negative $K_z$ in Eq. (\ref{effective}) raises 
the value of $\tilde{K}$ to $+\infty$, the ferromagnetic instability may occur, 
at which the velocity $v$ also vanishes. 
The other case is the {\it staggered} order of the imbalanced magnetization 
$\mu_j$; If $ 9/\tilde{K} < 2$, $\theta$-field is locked and then the staggered 
component of $\tau_j^x$ can have a finite expectation value through 
the relation 
$\hat P_j\mu_j\hat P_j = -\tau^x_j \sim (-)^j \cos (\sqrt{\pi}\theta)$.
Here, we note that, in the following numerical computations, 
the ferro-chirality oder actually appears, but a uniform order 
of $\mu_j$ is realized rather than the staggered type.

\section{numerical results}
\label{Sec3}

Now we apply DMRG to the spin tube model (\ref{tube}) to quantitatively 
examine the transitions and orderings with the help of results 
in Sec.~\ref{Sec2}. We fix $J=1$ in the following numerical calculations.

\subsection{chirality liquid phase}

\begin{figure}[ht]
\begin{center}\leavevmode
\includegraphics[width=0.9\linewidth,angle=0]{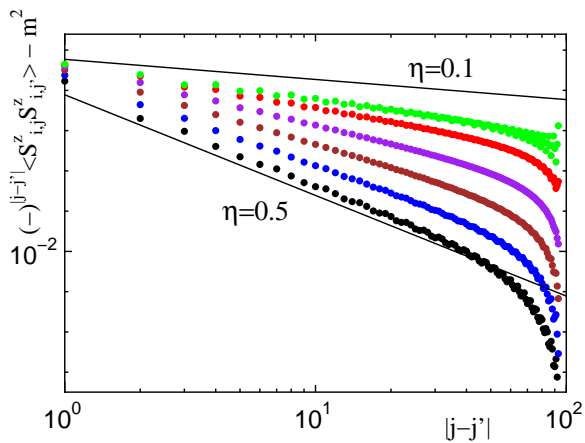}
\caption{(color online) Longitudinal spin correlation function 
$\langle S_{i,j}^{z}S^z_{i,j'} \rangle$ of the spin tube for 
$J'=0.01$, 0.1, 0.2, 0.3, 0.4, and 0.45 from bottom to top, where 
$m= \frac{1}{6}$. Two solid lines indicate guides for 
$\eta=0.5$ (XY chain) and 0.1.}
\label{fig:2}
\end{center}
\end{figure}

First, we focus on a sufficiently strong-rung coupling region. 
In Fig.~\ref{fig:2}, we present the longitudinal spin correlation function 
$\langle S^z_{i,j}S^z_{i,j'}\rangle $ for $L=96$ systems 
with $J'=0.01$, $\cdots$, 0.45. 
The rapid decay near the right edge in Fig.~\ref{fig:2} 
comes from the open boundary effect.
Thus it can be confirmed that the correlation function 
follows a power-law decay for $|j-j'|\alt 50$: 
$\langle S^z_{i,j}S^z_{i,j'}\rangle -m^2 \sim (-)^{j-j'}|j-j'|^{-\eta}$, 
where $m=\frac{1}{6}$ is the uniform magnetization per spin and 
$\eta$ is the critical exponent. This decay fashion is in agreement with 
the prediction from the effective TL-liquid theory~(\ref{TLL}). 
We can also see that $\eta$ becomes close to 0.5 in 
the $J'=0$ limit, where the Hamiltonian (\ref{effective}) reduces 
to the XY model. As $J'$ increases, $\eta$ approaches zero 
toward the ferro-chirality transition. 
Utilizing the effective field theory (\ref{TLL}) 
based on the XXZ chain (\ref{effective}), 
we can evaluate the critical exponent $\eta$ in the strong 
rung-coupling region $J\gg J'$. The value upto the second order of $J'$ 
is given by $\eta \simeq 0.5 - 0.885J' + 0.640J'^2 +\cdots$, 
where we have assumed 
the nonuniversal parameter $b_0=0.5424\cdots$~\cite{nonuniversal}.
We have confirmed that this value of $\eta$ is semi-quantitatively 
consistent with the numerically estimated value from 
the correlation function of Fig. \ref{fig:2} in $J\gg J'$.
From these results, we conclude that the gapless non-magnetic chirality 
excitation is described by the effective model (\ref{effective}).
Here, note that the width of the plateau is sufficiently large for $J'<0.5$ 
and the transverse correlator $\langle S^x_{i,j}S^x_{i,j'}\rangle$ 
exponentially decays, indicating that the magnetic excitation has a 
large gap corresponding to the plateau width.

\subsection{ordered phases}

\begin{figure}[bt]
\begin{center}\leavevmode
\includegraphics[width=0.9\linewidth,angle=0]{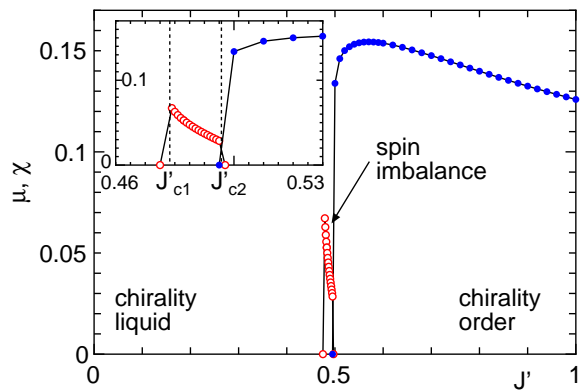}
\caption{(color online) Expectation value of the order parameters 
$\chi$ (solid circle) and $\mu$ (open circle). 
Inset shows these order parameters around the transition points. 
The vertical broken lines indicate the transition points 
$J'_{c1}$ and $J'_{c2}$.}
\label{fig:3}
\end{center}
\end{figure}

As $J'$ further increases, the negative $K^z$ derives the system 
toward a ferro-chirality ordered state with $\langle \chi_j\rangle\neq 0$.
Figure \ref{fig:3} illustrates the results of the order parameters $\chi=\langle\chi_j\rangle$ and $\mu=\langle\mu_j\rangle$. 
Here, $\chi$ is observed at the center triangle of the tube of size
 $L=120$~\cite{Okunishi08} and $\mu$ is the bulk expectation value 
based on the infinite system DMRG.
We have checked that the boundary effect is negligible within computations for $L=96, 120$ and $144$.
From the main panel, we can see two quantum phase transitions near $J'=0.5$. 
Note that the plateau width around $J'= 0.5$ is about $ 0.5J$, which is 
sufficiently larger than the energy scale of the non-magnetic 
chirality excitation.
Figure~\ref{fig:3} clearly shows the emergence of the ferro-chirality order 
in $J'>J'_{c2} = 0.496$, which is consistent with the 
effective model~(\ref{effective}). 
We have confirmed that this ferro-chirality order extends to $J'>1$ and 
thus it would be adiabatically connected to the vector chirality order 
in the region of the weakly-coupled three chains~\cite{Sato07}.
Here, note that both $\langle S^x_{i,j}S^x_{i,j'}\rangle$ and $\langle S^z_{i,j}S^z_{i,j'}\rangle$ show exponential decays in $J'>J'_{c2}$ and thus 
the magnetic and chirality excitations have finite gaps 
in this chirality ordered phase.

\begin{figure}[bt]
\begin{center}\leavevmode
\includegraphics[width=0.8\linewidth,angle=0]{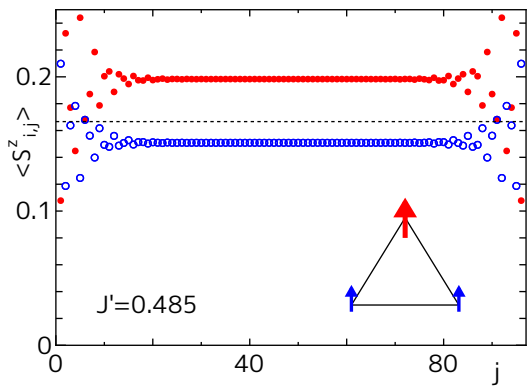}
\caption{(color online) Spin profile  $\langle S_{i,j}^z\rangle $ of the spin 
imbalance phase. The tube length is $L=96$ and the inter triangle coupling is 
$J'=0.485$. Solid circles denote the expectation value of the 
most polarized spin in each rung triangle and the open circles 
correspond to those of remainsing two spins on the triangle. 
The horizontal broken lines is the averaged 
magnetization of each rung in the plateau state.}
\label{fig:4}
\end{center}
\end{figure}

\begin{figure}[bt]
\begin{center}\leavevmode
\includegraphics[width=0.8\linewidth,angle=0]{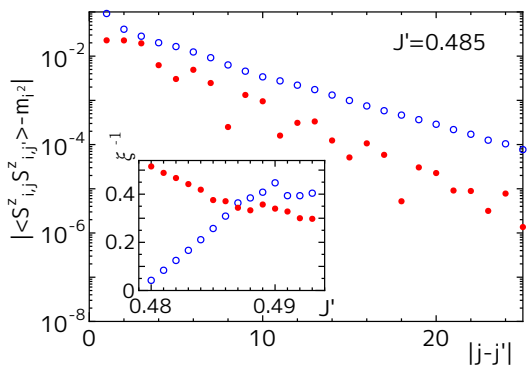}
\caption{(color online) Correlation function 
$|\langle S_{i,j}^z S^z_{i,j'}\rangle  -m_i^2 |$ in the spin imbalance phase. 
Solid circles is the correlator for the chain consisting of 
the most polarized spins on the unit triangules, 
and the open circles correspond to that for the remaining two chains.
Inset represents $J'$ dependence of the inverse correlation length 
$\xi^{-1}$ along the chains for the most and less polarized spins 
in the triangle.}
\label{fig:5}
\end{center}
\end{figure}

From the inset of Fig.~\ref{fig:3}, we also find that 
the spin imbalance phase emerges in a narrow region $J'_{c1}<J'<J'_{c2}$ 
with $J'_{c1} \simeq 0.478$.
In this region, the symmetry of the unit triangle reduces to the isosceles type, where the expectation value of one spin of each rung triangle is larger 
than those of the remaining two spins: 
$\langle S_{i,j}^z\rangle > \langle S_{i+1,j}^z\rangle =\langle S_{i+2,j}^z\rangle$. 
In Fig. \ref{fig:4}, we present the $\langle S^z_{i,j}\rangle$ distribution for $J'=0.485$, which exhibits a typical spin profile of the spin-imbalance state. 
The open-boundary effect rapidly decays and a uniform spin imbalance along the chain direction is realized around the center of the tube. 
Figure~\ref{fig:5} shows a semi-log plot of $\langle S^z_{i,j}S^z_{i,j'}\rangle-m_i^2$, where $m_i$ is the bulk expectation value of $S^z_{i,j}$ calculated at the center of the tube.
The exponential decay of the correlation functions in Fig.~\ref{fig:5} indicates that the system is gapful. 
We note that the imbalanced nature is present not only in the magnetization profile, but also in the spin correlation functions. 
As we see from the inset of Fig. \ref{fig:5}, the correlation length for the less polarized spins becomes divergent as $J' \to J'_{c1} + 0$, while that for the most polarized spin remains finite value.
This suggests that the instability of the spin imbalance toward the chirality 
liquid state ($J'<J'_{c1}$) may be governed by the fluctuation of the less polarized spins of the triangle, although the critical behavior of $\mu$ cannot be determined within the accuracy of the present DMRG results.
As $J'$ increases, the correlation lengths of the most polarized spin and the remaining two become comparable with each other and finally arrives at the ferro-chirality transition point $J'_{2c}$.
Here, we note that, for $0.485 \alt J'<J'_{c2}$, the spin correlation functions becomes highly oscillating and thus precise estimation of the correlation length is difficult.
We stress that this imbalanced order cannot be described by the effective model~(\ref{effective}). 
This suggests that the hybridization of $T^z=3/2$ sector plays an essential role in the imbalanced phase (see the following paragraphs).
On the other hand, the jump of the order parameters at $J'_{c2}$ clearly shows that the transition at $J'=J'_{c2}$ is of first order, where the two different symmetry breakings are switched.

Let us discuss the nature of the spin-imbalance phase in more detail.
As we disucssed above, the imbalanced order is uniform along the leg direction, while the field theory based on the effective model~(\ref{effective}) suggests the emergence of a staggered imbalance order 
($\langle\mu_j\rangle=-\langle\mu_{j+1}\rangle$).
This mismatch of the effective theory may be attributed to the fact that the imbalanced order is located at very vicinity of the ferro-chirality transition point $J'_{c2}$, where the velocity $v$ almost vanishes and thus the system becomes fragile.
Furthermore, we find that the rapid increase of $\mu$ in $J' > J'_{c1}$ causes a rapid raise of the energy of the unit triangle (DMRG data are not presented here), implying that the effect of $J'$ nonperturbatively reduces the energy of the intra-triangle bonds.
Thus it is suggested that the role of the intra-triancle coupling becomes essential and thus the $T=\frac{3}{2}$ sector certainly hybridizes into the plateau state in the spin-imbalance phase.

\begin{figure}[bt]
\begin{center}\leavevmode
\includegraphics[width=0.9\linewidth,angle=0]{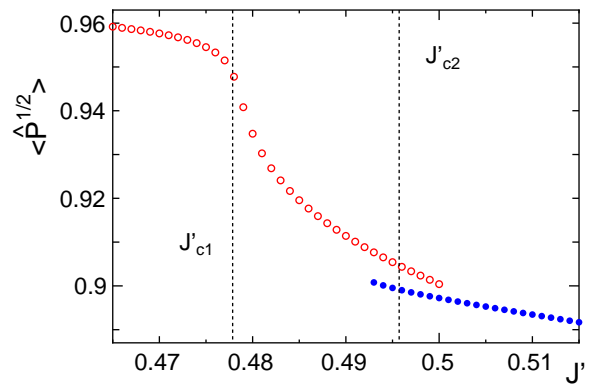}
\caption{(color online) Expectation value of the projection operator 
$\langle \hat{P}^{1/2} \rangle$ on the unit triangle.}
\label{fig:6}
\end{center}
\end{figure}

The effective model (\ref{effective}) is based on the massive weight of the $T=\frac{1}{2}$ sector, while the mixing of the $T=\frac{3}{2}$ sector is possibly essential for the spin imbalance phase. 
We should thus investigate the expectation value of $\hat{P}^{1/2}_j=({\bol T}_j^2/3 -5/4)$, which is the projection operator into the $T=\frac{1}{2}$ sector. 
In the $J'=0$ limit, $\langle \hat{P}^{1/2}_j \rangle=1$ and it gradually 
decreases up to $J'_{c1}$. 
Figure~\ref{fig:6} shows $\langle \hat{P}^{1/2}\rangle$ around the transition points, which is obtained by the infinite DMRG.
In the figure, we can see that the behavior of $\langle \hat{P}^{1/2}\rangle$ drastically changes at $J_{c1}\simeq 0.478$ and $J_{c2} =0.496$.
In $J_{c1}<J'<J_{c2}$, $\langle \hat{P}^{1/2} \rangle$ rapidly decreases with increasing $J'$. 
This supports that the driving mechanism of the spin-imbalance phase relies on the mixing of the $T=\frac{3}{2}$ sector.
Although a Berezinskii-Kosterlitz-Thouless (BKT) type transition accompanying the $Z_3$ symmetry breaking~\cite{nomura} is naively expected at $J_{c1}$, the nature of the phase transition might be essentially modified by the 
$T=\frac{3}{2}$ sector. 
However, we may claim within the present analysis that$\langle \hat{P}^{1/2} \rangle$ is continuously changed around $J_{c1}$, suggesting a continuous quantum phase transition. 
Further analysis is necessary to completely determine the nature of 
this transition, including the universality class. 
On the other hand, there exists a clear jump of $\langle \hat{P}^{1/2} \rangle$ at $J_{c2} =0.496$. 
The two branches near $J_{c2}$ represent two self-consistent solutions corresponding to the chirality ordered and spin-imbalance states in the DMRG iterations; 
the solution of the previous parameter is used as an initial state for the next parameter, so that the metastable states can be reproduced.
By comparing energies of the two branches, the first-order transition point can be determined as $J'_{c2} \simeq 0.496$.
This result is consistent with the behaviors of the order parameters in Fig.~\ref{fig:3}.

\section{Conclusions and discussions}
\label{Sec4}

In conclusion, we have explored the quantum phase transitions 
of the $\frac{1}{3}$ plateau state of the spin tube. 
In contrast to the usual plateaus of one-dimensional spin systems 
(chains and ladders), the chirality degree of freedom generated from 
the tube structure plays crucial roles. 
The results are summarized in Fig.~\ref{fig:Phases} (b), where the chirality 
liquid phase with gapless non-magnetic excitations, the spin-imbalance phase 
and the ferro-chirality phase emerge.
The qualitative features of these phases may be explained by the effective 
chirality model~(\ref{effective}) and the $S_3$-symmetry breakings. 
However, the precise analysis of the projection operator $\hat{P}^{1/2}$ 
has revealed that the uniform spin imbalance order is driven by mixing of 
the $T=\frac{3}{2}$ sector, which is beyond the scope of the effective 
model~(\ref{effective}). 
The transition between the chirality liquid and the spin-imbalance phase is 
of continuous type, and the fluctuation of less polarized spins 
in the imbalance phase becomes divergent near the transition. 
On the other hand, the transition between the spin-imbalance and 
ferro-chirality ordered phases is shown to be of first order type.

Here it should be commented that another spin-imbalance phase with gapless magnetic excitations is expected in a high magnetic field~\cite{SatoSakai07}. 
Its connection to the present spin-imbalance phase may be an interesting problem for through understanding of mechanisms of the spin imbalance. 
As we mentioned in the introduction, a chirality-ordered spin liquid appears in the weak rung-coupling region $J\ll J'$ in magnetic fields~\cite{Sato07}. 
This spin liquid is expected to change into the 1/3 plateau state with the chirality order~\cite{Sato07} via a BKT transition~\cite{cabra} at the order of $J/J'=0.1$. 
Combining our present results with this, we can conclude that, as $J'$ increases from the strong rung limit,  the chirality liquid, spin-imbalance order, ferro-chirality order, and ferro-chirality-ordered spin liquid can be observed at $m=\frac{1}{6}$ in order.

An important aspect of the spin tube is that the phase transitions occur without destroying the plateau. 
The energy scale of the chirality is significantly lower than the width of the large plateau.
Therefore, for example, a specific heat measurement will solely observe a linear temperature dependence originating from the chirality modes in the wide spin-gapped plateau region of $J'<J'_{c1}$, in contrast to the twisted tube~\cite{Nojiri2}.
From experimental viewpoint, moreover, another plausible feature of the spin tube is that the gapped chirality order is expanded in the wide range of $J'$, which is contrasted to the narrow chirality-ordered phases with gapped magnetic excitations in the classical XY model on triangular lattice~\cite{Miyashita} and spin-$S$ $J_1$-$J_2$ chains~\cite{Hikihara,Sato11}. 
If a coupling between chirality and electric polarization is introduced, the chirality order can induce a ferro-electric polarization in spite of the absence of any magnetic ordering. 
Also,  the similar chirality degree of freedom is discussed in the coupled trimer model, which may reduce to the spin tube in an anisotropic limit\cite{kamiya}.
We thus believe that the spin tube provides a fascinating play ground of 
the chirality degrees in the realistic experimental situation.

\begin{acknowledgments}

This work has been partly supported by Grants-in-Aid for Scientific Research 
(No.$\,$23340109, 23540442, 21740295, 23540388) and  Priority Area 
"Novel States of Matter Induced by Frustration" (No.22014012, 22014016) 
from MEXT, Japan. Numerical computations were partly performed at 
the Supercomputer Center, ISSP, University of Tokyo and the Computer Room, 
Yukawa Institute, Kyoto University.

\end{acknowledgments}


\begin{thebibliography}{99}

\bibitem{frustration}
For example, {\it Frustrated spin systems}, ed. H. T. Diep, 
(World Scientific, 2005).

\bibitem{Miyashita}
S. Miyashita and H. Shiba, J. Phys. Soc. Jpn. {\bf 53}, 1145 (1984). 

\bibitem{chirality} 
H. Kawamura, J. Phys. Condns. Matter. {\bf 10}, 4707 (1998).


\bibitem{Shannon}
N. Shannon, T. Momoi, and P. Sindzingre, Phys. Rev. Lett. {\bf 96}, 
027213 (2006). 

\bibitem{Okunishi08}
K. Okunishi, J. Phys. Soc. Jpn. {\bf 77}, 114004 (2008).

\bibitem{HKMF}
T. Hikihara, L. Kecke, T. Momoi, and A. Furusaki, 
Phys. Rev. B {\bf 78}, 144404 (2008). 

\bibitem{Furukawa}
S. Furukawa, M. Sato and S. Onoda, Phys. Rev. Lett. {\bf 105}, 257205 (2010). 


\bibitem{Wang}
K. F. Wang, J.-M. Liu and Z. F. Ren, Adv. Phys. {\bf 58}, 321 (2009). 


\bibitem{schulz}
H. J. Schulz, in {\it Correlated Fermions and Transport in Mesoscopic 
Systems}, eds. T. Martin, G. Montambaux, J. Tran Than Van (1996); 
cond-mat/9605075. 

\bibitem{kawano}
K. Kawano and M. Takahashi, 
J. Phys. Soc. Jpn. {\bf 66}, 4001 (1997). 


\bibitem{cabra}
D. C. Cabra, A. Honecker and P. Pujol, Phys. Rev. Lett. 
{\bf 79}, 5126 (1997); Phys. Rev. B {\bf 58}, 6241 (1998).
\bibitem{Sato07}
M. Sato, Phys. Rev. B {\bf 75}, 174407 (2007). 
\bibitem{SatoSakai07}
M. Sato and T. Sakai, Phys. Rev. B {\bf 75}, 014411 (2007).

\bibitem{mila}
A. Luscher, R. M. Noack, G. Misguich, V. N. Kotov and F. Mila, 
Phys. Rev. B {\bf 70}, 060405(R) (2004). 
\bibitem{sakai}
T. Sakai, M. Sato, K. Okunishi, Y. Otsuka, K. Okamoto, C. Itoi, Phys. Rev. B {\bf 78}, 184415 (2008). 
\bibitem{arikawa}
S. Nishimoto and M. Arikawa, Phys. Rev. B {\bf 78}, 054421 (2008).

\bibitem{review} T. Sakai, M. Sato, K. Okunishi, K. Okamoto, C. Itoi, 
J. Phys. Condens. Matter. {\bf 22}, 403201 (2010).



\bibitem{okunishi}
K. Okunishi, S. Yoshikawa, T. Sakai and S. Miyashita, Prog. Theor. Phys. Suppl. {\bf 159}, 297 (2005).

\bibitem{fouet}
J.-B. Fouet, A. L\"auchli, S. Pilgram, R. M. Noack, and F. Mila, Phys. Rev. B 
{\bf 73}, 014409 (2006). 

\bibitem{penc}M. Lajko, P. Sindzingre, and K. Penc,  Phys. Rev. Lett. {\bf 108}, 017205 (2012).



\bibitem{Nojiri}
J. Schnack, H. Nojiri, P. K\"ogerler, G. J. T. Cooper and L. Cronin, 
Phys. Rev. B {\bf 70} 174420 (2004).
\bibitem{Nojiri2}
N. B. Ivanov, J. Schnack, R. Schnalle, J. Richter, P. K\"ogerler, 
G. N. Newton, L. Cronin, Y. Oshima, and H. Nojiri, Phys. Rev. Lett. {\bf 105}, 
037206 (2010).

\bibitem{Manaka}
H. Manaka, Y. Hirai, Y. Hachigo, M. Mitsunaga,M. Ito, and N. Terada, 
J. Phys. Soc. Jpn {\bf 78}, 093701 (2009). 
\bibitem{Manaka2}
H. Manaka, T. Etoh, Y. Honda, N. Iwashita, K. Ogata, N. Terada, T. Hisamatsu, 
M. Ito, Y. Narumi, A. Kondo, K. Kindo, and Y. Miura, 
J. Phys. Soc. Jpn. {\bf 80}, 084714  (2011).






\bibitem{Giamarchi}
See, for example, T. Giamarchi, {\it Quantum Physics in One Dimension} 
(Oxford University Press, 2004). 


\bibitem{nonuniversal}
T. Hikihara and A. Furusaki, Phys. Rev. B {\bf 69}, 064427 (2004). 


\bibitem{nomura}
K. Nomura, J. Phys. A: Math. Gen. {\bf 28}, 5451 (1995).





\bibitem{Hikihara}
T. Hikihara, M. Kaburagi, and H. Kawamura, 
Phys. Rev. B {\bf 63}, 174430 (2001). 

\bibitem{Sato11}
M. Sato, S. Furukawa, S. Onoda, and A. Furusaki, Mod. Phys. Lett. 
B {\bf 25}, 901 (2011); arXiv:1101.1374; S. Furukawa, M. Sato, 
S. Onoda, and A. Furusaki, in preparation. 

\bibitem{kamiya} Y. Kamiya and C. D. Batista, arXiv:1110.4120.

\end{thebibliography}

\end{document}